\documentclass[11pt]{article}

\usepackage[a4paper,top=2.5cm,bottom=2.cm,left=2.8cm,right=2.5cm,marginparwidth=1.75cm]{geometry}

\usepackage{authblk}
\usepackage{lineno}
\usepackage{amssymb}
\usepackage{amsmath}
\usepackage{amsthm}
\usepackage{epsfig}
\usepackage{graphicx}
\usepackage{graphics}
\usepackage{float}
\usepackage{multirow}
\usepackage{color}
\usepackage{lineno}
\usepackage[normalem]{ulem} 
\usepackage{makeidx}
\usepackage{xspace}
\usepackage{cite}
\usepackage[colorlinks=true, linktoc=page, linkcolor=blue, citecolor=blue, urlcolor=blue]{hyperref}
\usepackage{caption} 
\begin{document}

\renewcommand\thesection{\arabic{section}}
\title{Test of quantum nonlocality via vector meson decays to $K_SK_S$}

\author[*]{Pei-Cheng Jiang\thanks{jiangpc@stu.pku.edu.cn}}
\author[*]{Xuan Wang\thanks{wangxuan15@pku.edu.cn}}
\author[*]{Da-Yong Wang\thanks{dayong.wang@pku.edu.cn}}
\affil[*]{School of Physics and State Key Laboratory of Nuclear Physics and Technology

Peking University, Beijing 100871, China}
\renewcommand*{\Affilfont}{\small\it}
\renewcommand\Authands{ and }

\date{}
\maketitle

\vspace{-1cm}

\begin{abstract}
In the system of a pair of quantum-entangled neutral kaons from meson decays, when one kaon collapses into the $K_S$ state, the other will collapse instantaneously into the $K_L$ state due to entanglement and nonlocality. However, if the alternative hypothesis is correct and there's a time window during which one kaon is unaware that the other has decayed, some quantum mechanically prohibited $K_SK_S$ decays may occur. We calculate the branching ratios of $K_SK_S$ in vector meson decays under the locality hypothesis and compare them with experimental results. We show that the branching ratio of $J/\psi\rightarrow K_SK_S$ under locality assumption is already excluded by the BESIII experimental upper limit. Additional experimental results are proposed to perform this test in the future.
\end{abstract}

\section{Introduction}
In 1935, Einstein, Podolsky, and Rosen (EPR) posed the question of whether or not quantum mechanics offers a complete description of reality \cite{1935Phys}. They assumed the locality principle, which states that interference effects should travel at the speed of light or slower between two objects. While in quantum mechanics, the measurement of one particle in an entangled system has an instantaneous effect on the other due to explicit nonlocality. Quantum-mechanics nonlocality tests have been extensively carried out in optics and atomic physics studies \cite{freedman,thompson,aspect}. All of the results are consistent with quantum mechanical predictions. High-energy physics measurements may also reveal the incompatibility of quantum physics with local realism \cite{li_new_2009}.

Noninstantaneous interaction in the neutral kaon system is sensitive to testing the nonlocality of quantum mechanics. Under Einstein's assumption of locality, the neutral kaon system must produce some $K_SK_S$ decays in the space-like region, despite the fact that quantum mechanics prohibits this process \cite{eberhard_testing_1993}. 

In this paper, we calculate the branching ratios of vector meson decays to $K_SK_S$ under the locality assumption and compare them with experimental results in order to test for nonlocal phenomena in neutral kaon systems. Additional experimental measurements of such channels are proposed to perform this test in the future.

\section{Entangled neutral kaons system}
We discuss the test in the reaction 
\begin{equation}
V\rightarrow K^0\bar{K}^0,
\end{equation}
where V is a vector meson ($\phi$, $J/\psi$, $\Upsilon$...) with quantum numbers $J^{PC}=1^{--}$. For the entangled system of two neutral kaons, immediately after the decay (at time zero), the quantum–mechanical state could be depicted as

\begin{equation}
\begin{aligned}
|\phi(0)\rangle=\frac{1}{\sqrt{2}}\left\{\left|K^{0}\right\rangle_{a}\left|\bar{K}^0\right\rangle_{b}-\left|\bar{K}^0\right\rangle_{a}\left|K^{0}\right\rangle_{b}\right\}
\\
=\frac{1}{\sqrt{2}}\left\{\left|K_{S}\right\rangle_{a}\left|K_{L}\right\rangle_{b}-\left|K_{L}\right\rangle_{a}\left|K_{S}\right\rangle_{b}\right\},
\end{aligned}
\end{equation}

where \textbf{a} and \textbf{b} denote the two kaons' opposing directions of motion. When the effect of CP violation is ignored, the CP eigenstates are identical to $K_S$ and $K_L$, which are short-lived and long-lived neutral kaons, respectively. Thus there is no $K_SK_S$ component in the decay products.

The time evolution of states $K_S$ and $K_L$ is given by 
\begin{equation}
\left|K_{S}(t)\right\rangle=\left|K_{S}\right\rangle \exp \left(-\alpha_{s} t\right), \quad\left|K_{L}(t)\right\rangle=\left|K_{L}\right\rangle \exp \left(-\alpha_{L} t\right)
\end{equation}
respectively, where $t$ is the particle proper time and 
\begin{equation}
\alpha_{s}=\frac{1}{2} \Gamma_{s}+i m_{s}, \quad \alpha_{L}=\frac{1}{2} \Gamma_{L}+i m_{L}.
\end{equation}

In Eq. (4), $\Gamma_S$ ($\Gamma_L$) and $m_S$ ($m_L$) are the decay rates and masses for $K_S$ ($K_L)$, respectively. According to quantum mechanics, the decay amplitude of the two kaons' states into final states $f_a$ and $f_b$ at proper times $t_a$ and $t_b$ can be written as \cite{amplitude}:
\begin{equation}
\begin{aligned}
A\left(f_{a}, t_{a} ; f_{b}, t_{b}\right)=&\frac{1}{\sqrt{2}}[\left\langle f_{a}|T| K_{S}\left(t_{a}\right)\right\rangle\left\langle f_{b}|T| K_{L}\left(t_{b}\right)\right\rangle
\\
&-\left\langle f_{a}|T| K_{L}\left(t_{a}\right)\right\rangle\left\langle f_{b}|T| K_{S}\left(t_{b}\right)\right\rangle]
\\
=&\frac{1}{\sqrt{2}}[\left\langle f_{a}|T| K_{S}\right\rangle\left\langle f_{b}|T| K_{L}\right\rangle e^{-i \alpha_{S} t_{a}} e^{-i \alpha_{L} t_{b}}
\\
&-\left\langle f_{a}|T| K_{L}\right\rangle\left\langle f_{b}|T| K_{S}\right\rangle e^{-i \alpha_{L} t_{a}} e^{-i \alpha_{S} t_{b}}],
\end{aligned}
\end{equation}
where $T$ is the transition operator from the two kaons' states to the final states.

Up to the moment of the first kaon decay, the two kaons are entangled and the decay rate for $K^0\bar{K}^0\rightarrow \rm anything$ can be computed from equation (5), using the definition of (3) and (4):
\begin{equation}
\begin{aligned}
\Gamma_{\rm ent}\left(t_{a}, t_{b}\right) =&N \Sigma_{f_{a}, f_{b}}\left|A\left(f_{a}, t_{a} ; f_{b}, t_{b}\right)\right|^{2} \\
=&\frac{N}{2} \Gamma_{L} \Gamma_{\mathrm{S}}\{e^{-\Gamma_{S} t_{a}-\Gamma_{L} t_{b}}+e^{-\Gamma_{S} t_{b}-\Gamma_{L} t_{a}}\\
&-2 \cos \left[\Delta\left(t_{b}-t_{a}\right)\right] e^{-\frac{1}{2}\left(t_{b}+t_{a}\right)\left(\Gamma_{S}+\Gamma_{L}\right)}\},
\end{aligned}
\end{equation}
where $\Delta = m_L-m_S$ and normalization factor N $\approx 1+2\left(\Gamma_{L} / \Gamma_{S}\right)\approx 1.0035$ guarantees the integral of $\Gamma_{\rm ent}$ to be 1.

After the decay of the first kaon, the quantum interference between the two kaons disappears and the decay rate becomes
\begin{equation}
\Gamma_{\rm non\_ent}\left(t_{a}, t_{b}\right)=\frac{1}{2} \Gamma_{L} \Gamma_{\mathrm{S}}\left\{e^{-\Gamma_{S} t_{a}-\Gamma_{L} t_{b}}+e^{-\Gamma_{S} t_{b}-\Gamma_{L} t_{a}}\right\}.
\end{equation}

The decay rate at time $t_a$ of kaon \textbf{a} can be expressed as 
\begin{equation}
\Gamma_{a}\left(t_{a}\right)=\int_{0}^{t_{a}} d t_{b} \Gamma_{\rm non\_ent}\left(t_{a}, t_{b}\right)+\int_{t_{a}}^{+\infty} d t_{b} \Gamma_{\rm ent}\left(t_{a}, t_{b}\right).
\end{equation}

Similarly, the decay rate at time $t_b$ of kaon \textbf{b} is given by Eq. (8) with the replacement $\textbf{a} \leftrightarrow \textbf{b}$. 

\begin{figure}[h] 
\centering
\includegraphics[width=0.7\textwidth]{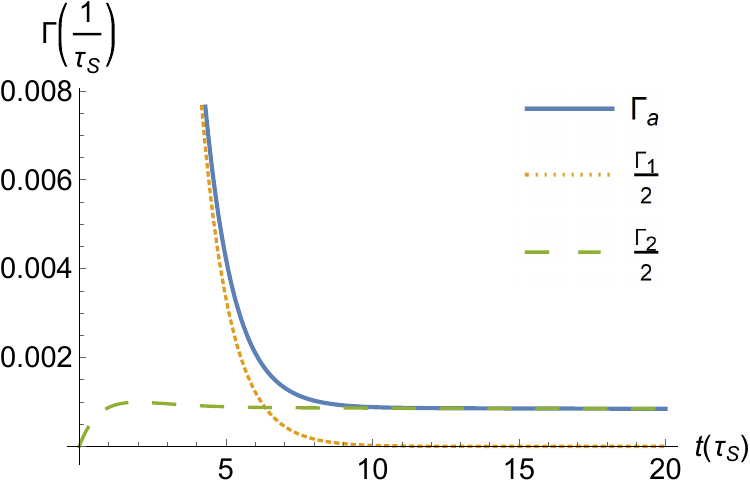} 
\caption{$\Gamma_a$ (solid blue line) is the decay rate of kaon \textbf{a} at time t in units of $K_S$ life time $\tau_S$ in its own rest frame. $\frac{\Gamma_1}{2}$ (orange dashed line) and $\frac{\Gamma_2}{2}$ (green dashed line) represent the decay rates of kaon \textbf{a} decaying first and second, respectively.} 
\label{Fig1}
\end{figure}

Combining the contributions of kaon \textbf{a} and kaon \textbf{b}, the decay rate of the first decay at time $t_1$ is
\begin{equation}
\Gamma_{1}(t_{1})=2 \int_{t_{1}}^{+\infty} d t_{2} \Gamma_{\rm ent}\left(t_{1}, t_{2}\right),
\end{equation}
and the decay rate of the second decay at time $t_2$ is
\begin{equation}
\Gamma_{2}(t_{2})=2 \int_{0}^{t_{2}} d t_{1} \Gamma_{\rm non\_ent}\left(t_{1}, t_{2}\right).
\end{equation}
Since the sequence of \textbf{a} and \textbf{b} decays can be random, $\Gamma_a(t) = \Gamma_1(t)/2+\Gamma_2(t)/2$. The decay rates of $\Gamma_a(t)$, $\Gamma_1(t)/2$ and $\Gamma_2(t)/2$ as a function of $K_S$ life time $\tau_S$ are shown in Fig. \ref{Fig1}.

There are three different schemes to describe the whole decay process.
The first is the quantum-mechanics interpretation. At the beginning, both kaons have a decay rate of $\Gamma_1(t)/2$. When one kaon (whether \textbf{a} or \textbf{b}) decays first at time $t_1$, the decay rate of the other kaon changes instantaneously from $\Gamma_1(t)/2$ to $\Gamma_2(t)/2$.
The second description comes from the hidden-variable theory. The two kaons are determined to be $K_S$ or $K_L$ with certain decay rates, and their final products are identical to those of quantum mechanics.
The third one is the locality assumption to test. When one kaon decays first at time $t_1$, the other kaon is unaware that the first decay kaon has decayed until the information propagating with the velocity c arrives. And until $t_1'=\gamma't_1$, where $\gamma^{\prime}=(1+\beta) /(1-\beta)$ ($\beta = v/c$, where v is the velocity of the kaons in the lab frame), the decay rate of the second decay kaon remains $\Gamma_1(t)/2$. Only at time $\gamma't_1$ does it change discontinuously to $\Gamma_2(t)/2$ \cite{roos_test_1980}.

Under the locality assumption, there is a lapse of time during which the second decay kaon may decay, not knowing whether the first kaon has decayed, thus it can not be influenced. At lab time $t_2$ , the second decay kaon could have been influenced by the decay of the first kaon only if it occurred at time $t_2/\gamma'$ or earlier. 

\begin{figure}[h] 
\centering 
\includegraphics[width=0.7\textwidth]{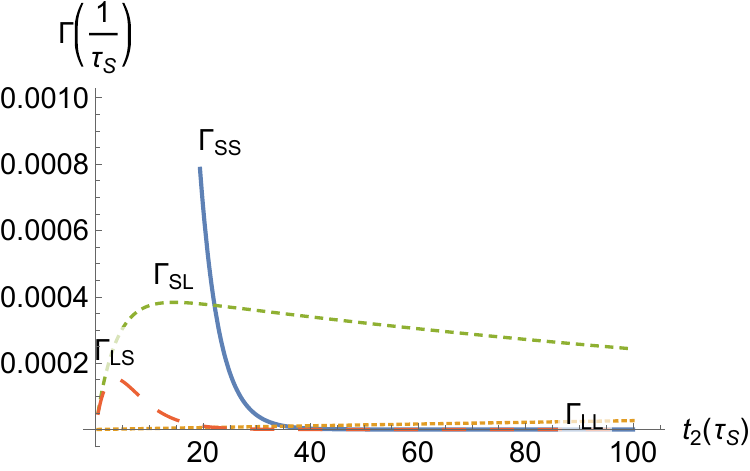} 
\caption{During the time window, the second decay kaon has no idea that the first kaon has decayed. Then the incoherent states $\left|K_{S}\right\rangle_{1}\left|K_{S}\right\rangle_{2}$, $\left|K_{S}\right\rangle_{1}\left|K_{L}\right\rangle_{2}$, $\left|K_{L}\right\rangle_{1}\left|K_{S}\right\rangle_{2}$, and $\left|K_{L}\right\rangle_{1}\left|K_{L}\right\rangle_{2}$ exist. Relative decay rates $\Gamma_{SS}$ (blue solid line), $\Gamma_{SL}$ (green dashed line), $\Gamma_{LS}$ (orange dashed line), $\Gamma_{LL}$ (yellow dashed line) in the lab frame are shown in this figure. It is necessary to mention that $\Gamma_{LL}$ will be dominant when $t_2>400\,\tau_S$.} 
\label{Fig2}
\end{figure}

As mentioned above, in the time window ($t_2/\gamma'$, $t_2$), the parity is not conserved and the $K^0\bar{K}^0$ state consists of the incoherent states: $\left|K_{S}\right\rangle_{1}\left|K_{S}\right\rangle_{2}$, $\left|K_{S}\right\rangle_{1}\left|K_{L}\right\rangle_{2}$, $\left|K_{L}\right\rangle_{1}\left|K_{S}\right\rangle_{2}$, and $\left|K_{L}\right\rangle_{1}\left|K_{L}\right\rangle_{2}$ with equal weights. The relative decay rates as a function of $K_S$ life time $\tau_S$ are shown in Fig. \ref{Fig2}. 

\section{Experimental effects}
To compare the branching ratios of $V\rightarrow K_SK_S$ under the locality assumption with experimental results, certain experimental effects such as CP violation and kaon regeneration need to be taken into account.

Only considering the $K_S$ decay mode $K_{S}\rightarrow \pi^{+} \pi^{-}$, the final state of reaction $V\rightarrow K_SK_S$ is $\pi^{+} \pi^{-}\pi^{+} \pi^{-}$.
Due to CP violation, the decay $K_L\rightarrow \pi^+\pi^-$ may happen, whose branching ratio is \cite{ptaa104} 
\begin{equation}
\zeta=\frac{\Gamma_S^0}{\Gamma_L}|\eta_{+-}|^2=1.967\times10^{-3},
\end{equation}
where $|\eta_{+-}|$ is the CP-violating amplitude ratio and $\Gamma_S^0$ represents the partial width of $K_S$ decaying into the $\pi^+\pi^-$ mode.
Then the states $\left|K_{S}\right\rangle_{1}\left|K_{L}\right\rangle_{2}$, $\left|K_{L}\right\rangle_{1}\left|K_{S}\right\rangle_{2}$, and $\left|K_{L}\right\rangle_{1}\left|K_{L}\right\rangle_{2}$ can be misreconstructed as $\left|K_{S}\right\rangle_{1}\left|K_{S}\right\rangle_{2}$. Therefore the branching ratio of $V\rightarrow K_SK_S$ should be corrected with (1+$\sigma$), where
\begin{equation}
\sigma =\frac{\int_0^{+\infty}dt_2[\zeta \Gamma_{SL}(t_2)+\zeta\Gamma_{LS}(t_2)+\zeta ^2\Gamma_{LL}(t_2)]}{\int_0^{+\infty} dt_2\Gamma_{SS}(t_2)}.
\end{equation}
In Eq. (12), we have 
\begin{equation}
\Gamma_{SS}(t_2)=\left(\frac{\Gamma_{S}^{0}} {\gamma}\right)^{2} \int_{F_1} d t_{1} \exp \left[\frac{-\Gamma_{S}\left(t_{1}+t_{2}\right)}{ \gamma}\right],
\end{equation}
where $\gamma$ is the Lorentzian factor $\gamma=1/\sqrt{1-\beta^2}$ and $F_1$ represents time interval $[t_{2} / \gamma^{\prime}  < t_{1} < t_{2}; 0 < t_{1}<+\infty; 0 < t_{2}<+\infty]$. Expressions of the others are analogously defined as that of $\Gamma_{SS}(t_2)$. The numerical values of $\sigma$ for different vector mesons are listed in Table 1. 

In the vacuum, $K_L$ and $K_S$ are eigenstates of the Hamiltonian. If a kaon is a $K_L$ or a $K_S$, it remains a $K_L$ or a $K_S$ till it decays or interacts in the detector \cite{eberhard_testing_1993}. While in the material, the different elastic cross sections for $K_0$ and $\bar{K}^0$ will change the phase relations between the $K_0$ and $\bar{K}^0$ \cite{cern_k0k0bar_1967} due to the fact that the $K_0$ meson interacts differently with matter (generally with protons and neutrons) than $\bar{K}^0$. To analyze the $V\rightarrow K_SK_S$ process, one also needs to estimate the kaon regeneration probability in the corresponding momentum range.
There are two kinds of regeneration processes: coherent regeneration, which appears in the strictly forward direction; and incoherent regeneration, which is elastic scattering on nuclei with incoherent addition of amplitudes. The latter could be mostly excluded during the data analysis utilizing the angle $\theta$ between the directions of incident and outgoing kaons. 
Knowing the difference $\Delta f$ between forward scattering amplitudes of $K^0$ and $\bar{K}^0$ by the atoms, the mean lifetime $\tau_S$ of the $K_S$, the kaon mass m, the $K_L-K_S$ mass difference $\Delta m$, and the time $t$ taken by the kaon in its own rest frame to traverse the material, one can predict the probability $p_{\rm regen}$ for a kaon to be coherently regenerated \cite{cern_k0k0bar_1967}: 
\begin{equation}
p_{\rm regen}=|\rho|^{2} p_{\rm thru},
\end{equation}
where $p_{\rm thru}=e^{-\nu \ell \sigma_{\rm tot}}$ ($\sigma_{\rm tot}$ is the total cross section to $K_L$ and the atoms of the material, $\nu$ is the atomic density, and $\ell$ is the thickness of the material) is the probability for a $K_L$ not to interact in the material,
$
\rho=\frac{\pi \nu}{\frac{1}{2 \tau_{S}}-i \Delta m} \frac{\Delta f}{m} \kappa
$,
and $\kappa=1-e^{\left(-\frac{1}{2 \tau_{S}}+i \Delta m\right) t}$.
Taking the BESIII experiment as an example, the $K_L\rightarrow K_S$ regeneration can happen in the beam pipe and the inner wall of the main draft chamber (MDC). If $K_S$ decays before entering the material in the detector, then $K_L$ will cross the material as a free particle \cite{hai-bo_about_2009}. In this case, the $K_SK_S$ will be generated due to the regeneration effect. According to the design report of BESIII \cite{ablikim_design_2010}, the beam pipe is 1.4 mm of Beryllium, at the radius of 32 mm away from the beam axis. The inner wall of MDC is 1.2 mm thick carbon fiber, with a radius of 59 mm. The overall probabilities of $K_0\bar{K}^0\rightarrow K_SK_S$ due to kaon regeneration are listed in Table 1.

\section{The calculation}
During the time window ($t_2/\gamma'$, $t_2$) in the lab frame the reaction will yield double $K_S$ events from the state $\left|K_{S}\right\rangle_{1}\left|K_{S}\right\rangle_{2}$ with probability $P_W(K_SK_S)$ and single $K_S$ events
from the $\left|K_{S}\right\rangle_{1}\left|K_{L}\right\rangle_{2}$ and $\left|K_{L}\right\rangle_{1}\left|K_{S}\right\rangle_{2}$ states with probability $P_W(K_SK_L)$. In addition, the experiment will detect single $K_S$ events outside the time window, with a probability of $P_{QM}(K_SK_L)$. The ratio of double $K_S$ events to single $K_S$ events will be given by
\begin{equation}
R=P_{W}\left(K_{S} K_{S}\right) /\left[P_{Q M}\left(K_{S} K_{L}\right)+P_{W}\left(K_{S} K_{L}\right)\right],
\end{equation}
where $P_{QM}(K_SK_L)$ is obtained by integrating $\frac{2}{\gamma^{2}} \frac{\Gamma_{S}^{0}}{\Gamma_{S}} \Gamma_{\rm non\_ent}\left(\frac{t_{1}}{\gamma}, \frac{t_{2}}{\gamma}\right)$ over the time-like fiducial region $F_2$\,$[0 < t_{1} < t_{2} / \gamma^{\prime} ; 0< t_{1}<+\infty;0 < t_2 < +\infty]$. 

At fixed time $t_2$ the probability of the first decay occurring in the space-like region $P_W$ can be obtained by integrating $\frac{2}{\gamma^{2}} \frac{\Gamma_{S}^{0}}{\Gamma_{S}} \Gamma_{\rm non\_ent}\left(\frac{t_{1}}{\gamma}, \frac{t_{2}}{\gamma}\right)$ in the time window $F_1$\,$[t_{2} / \gamma^{\prime} < t_{1} < t_{2};0 < t_{1}<+\infty;0< t_2 <+\infty]$. 

\begin{figure}[h]
\centering 
\includegraphics[width=0.7\textwidth]{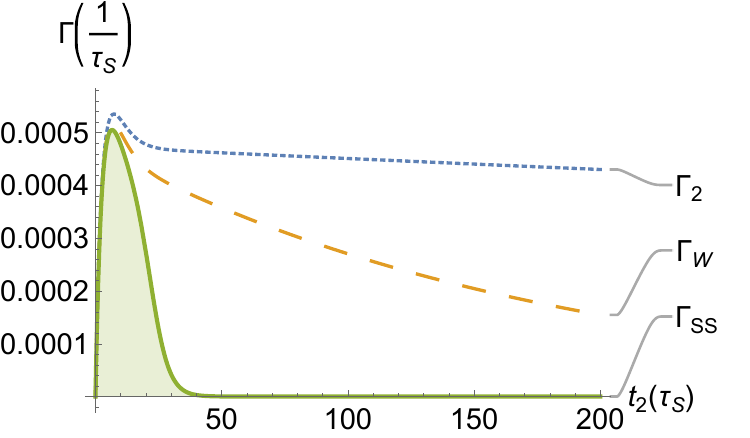} 
\caption{$\Gamma_2$ (blue dashed line) is the total decay rate at the second decay time $t_2$. $\Gamma_W$ (orange dashed line) represents the decay rate in the time window. The green shaded part $\Gamma_{SS}$ is the decay rate of $J/\psi\rightarrow K_SK_S$ ($K_S\rightarrow\pi^+\pi^-$) under the locality assumption.} 
\label{Fig3} 
\end{figure}

Next we can get the fraction of events decaying in the fiducial region as $K_SK_S$ is
\begin{equation}
p_{ss}=\frac{\int_0^{+\infty}dt_2\Gamma_{SS}(t_2)}{\int_0^{+\infty}dt_2[\Gamma_{SS}(t_2)+\Gamma_{SL}(t_2)+\Gamma_{LS}(t_2)+\Gamma_{LL}(t_2)]}.
\end{equation}
Multiply it by $P_W$ to get $P_W(K_SK_S)$ and the same with $P_W(K_SK_L)$. The relative decay rate of double $K_S$ events is shown in Fig. \ref{Fig3}. 

Considering the effects mentioned above, the corrected value of $R$ can be expressed as
\begin{equation}
R'=(1+\sigma)R+p_{\rm regen}.
\end{equation}
Knowing the branching ratio of $V\rightarrow K_SK_L$, multiplied by $R'$ and divided by $\Gamma_S^0/\Gamma_S$, the branching ratio of decay of $V\rightarrow K_SK_S$ can be obtained.

\section{Results and comparison with experiments}

For different vector mesons, the calculated results under the locality assumption are shown in Table 1. The uncertainty of $R'$ value mainly comes from the uncertainty of $K_L$'s decay rate. The branching ratios of $V\rightarrow K_SK_S$ expected under the locality assumption could be obtained with the branching ratios of $V\rightarrow K_SK_L$.

\begin{table}[t]
\caption{The ratio of double $K_S$ events to single $K_S$ events in reaction $V\rightarrow K^0\bar{K}^0$ and calculated branching ratio of $V\rightarrow K_SK_S$ under the locality assumption. $\sigma$ and $p_{\rm regen}$ are correction parameters from CP violation and kaon regeneration, respectively. $Br(\Upsilon(nS)\rightarrow K_SK_S)$ are not given because there are no accurate measurement results of $Br(\Upsilon(nS)\rightarrow K_SK_L)$.}
{\begin{tabular}{@{}cccccc@{}} \hline
Vector meson& Mass(MeV) &$\sigma$&$p_{\rm regen}$& $R'$ &$Br(V\rightarrow K_SK_S)$\\ \hline
$\phi(1020)$& $1019.46\pm0.016$ &$5.0\times10^{-5}$&$1.4\times10^{-6}$&$0.0005\pm0.0033$&$(0.4\pm2.5)\times10^{-2}$\\
$J/\psi$& $3096.90\pm0.006$ &$3.7\times10^{-4}$&$1.8\times10^{-6}$&$0.0196\pm0.0032$&$(5.5\pm1.0)\times10^{-6}$ \\
$\psi(2S)$& $3686.10\pm0.06$ &$5.1\times10^{-4}$&$1.8\times10^{-6}$&$0.0275\pm0.0032$&$(2.1\pm0.3)\times10^{-6}$\\
$\Upsilon(1S)$& $9460.3\pm0.26$ &$2.2\times10^{-3}$&$1.9\times10^{-6}$&$0.1282\pm0.0033$&-\\
$\Upsilon(2S)$& $10023.26\pm0.31$ &$2.3\times10^{-3}$&$1.9\times10^{-6}$&$0.1378\pm0.0033$&-\\
$\Upsilon(3S)$& $10355.2\pm0.5$ &$2.4\times10^{-3}$&$1.9\times10^{-6}$&$0.1425\pm0.0035$&-\\ \hline
\end{tabular}}
\end{table}

In the case of $\phi(1020)$, because its mass is small, the velocity of kaons is low and the time window is limited. The value of $R'$ is subject to considerable uncertainty, and thus performing this test with $\phi(1020)$ meson is difficult.

The calculated branching ratio of $J/\psi\rightarrow K_SK_S$ under the locality assumption is $(5.5\pm1.0)\times10^{-6}$ while the BESIII Collaboration has given an upper limit at 95\% C.L. $Br(J/\psi\rightarrow K_SK_S)<1.4\times10^{-8}$ in 2017 \cite{besiii_collaboration_branching_2017}. The upper limit is two orders of magnitude smaller than the expected value under the locality assumption, showing a significant violation of the theory. From the upper limit, the lower limit of the information transmission speed required by the locality hypothesis is 45.1 times the speed of light at the 95\% confidence level.

For $\psi(2S)$, the BES Collaboration has given an upper limit at 95\% C.L. $Br(\psi(2S)\rightarrow K_SK_S)<4.6\times10^{-6}$ in 2004 \cite{bai_search_2004} while the calculated value under the locality assumption is $(2.1\pm0.3)\times10^{-6}$. The experimental upper limit is still above the calculated value and unable to exclude the hypothesis of locality yet. With greater statistics, BESIII has the potential to verify or exclude the locality with $\psi(2S)$ decays. We encourage the BESIII colleagues to update this result in the near future.

For heavier vector mesons as listed in Table 1, the branching ratios of $V\rightarrow K_SK_L$ have not been measured yet. Since the $R'$ value increases as energy increases, the distinction between quantum mechanics and the locality hypothesis may be more pronounced. We propose the Belle-II and LHCb collaborations to pursue these studies to provide a conclusive test in the states of $\Upsilon(nS)$.

\section{Summary}
Under the locality assumption, we have estimated the branching ratios of $V\rightarrow K_SK_S$ and compared them with experimental upper limits. The result of $J/\psi$ is significantly less than the prediction of the locality hypothesis, indicating a preference for quantum mechanics' nonlocality. And the present upper limit of $\psi(2S)$ is compatible with the locality hypothesis' prediction. We encourage BESIII to perform this test in the near future with higher sensitivity. In addition, $\Upsilon(nS)$ could also be used to test the locality hypothesis, and we propose Belle-II and LHCb experiments to perform such studies.

\end{document}